\pdfoutput=1
\documentclass[10pt,journal,compsoc]{IEEEtran}

\ifCLASSOPTIONcompsoc
  \usepackage[nocompress]{cite}
\else
  \usepackage{cite}
\fi

\usepackage{amsmath,amssymb,amsfonts}
\usepackage{algorithmic}
\usepackage[ruled,vlined,linesnumbered]{algorithm2e}
\usepackage{graphicx}
\usepackage{textcomp}
\usepackage{xcolor}
\usepackage{url}
\usepackage{array}
\newcolumntype{M}{>{\centering\arraybackslash}m{5.0cm}}
\newcolumntype{N}{>{\centering\arraybackslash}m{3.0cm}}
\newcolumntype{J}{>{\centering\arraybackslash}m{2.0cm}}
\newcolumntype{K}{>{\centering\arraybackslash}m{1.5cm}}
\newcolumntype{L}{>{\centering\arraybackslash}m{1.0cm}}
\usepackage{multirow}
\usepackage{amssymb}

\usepackage{comment}
\usepackage{multirow}
\usepackage{listings}
\usepackage[shortlabels]{enumitem}
\usepackage{subcaption}
\usepackage{ifthen}
\usepackage[most]{tcolorbox}
\usepackage{alltt}
\usepackage{boxedminipage}
\usepackage{tikz}
\usepackage{balance}
\def\checkmark{\tikz\fill[scale=0.4](0,.35) -- (.25,0) -- (1,.7) -- (.25,.15) -- cycle;} 

\definecolor{red}{HTML}{9B0000}
\definecolor{lightred}{HTML}{FF5131}
\definecolor{green}{HTML}{006400}
\definecolor{lightgreen}{HTML}{9CFF57}
\definecolor{purple}{HTML}{7200CA}
\definecolor{verylightgrey}{HTML}{F1F1F1}

\definecolor{diffstart}{named}{blue}
\definecolor{diffincl}{named}{green}
\definecolor{diffrem}{named}{red}

\usepackage{listings}
  \lstdefinelanguage{diff}{
	basicstyle=\ttfamily\extrabold\scriptsize,
	morecomment=[f][\color{diffstart}]{@},
	morecomment=[f][\color{diffincl}]{+},
	morecomment=[f][\color{diffrem}]{-},
        keepspaces=true,
	identifierstyle=\color{black},
  }

\newboolean{showcomments}
\setboolean{showcomments}{true}
\ifthenelse{\boolean{showcomments}}
{ \newcommand{\mynote}[2]{\textcolor{red}{
			\fbox{\bfseries\sffamily\scriptsize#1}
			{\small$\blacktriangleright$\textsf{\emph{#2}}$\blacktriangleleft$}}}}
{ \newcommand{\mynote}[2]{}}

\ifthenelse{\boolean{showcomments}}
{ \newcommand{\hnote}[2]{\textcolor{blue}{
			\fbox{\bfseries\sffamily\scriptsize#1}
			{\small$\blacktriangleright$\textsf{\emph{#2}}$\blacktriangleleft$}}}}
{ \newcommand{\hnote}[2]{}}

\newcommand{\makeunderscoreletter}{\catcode`\_=11}

\usepackage{syntax}

\def \toolname {MLCatchUp}

\def \numcasestudy {112}

\begin{document}
\makeunderscoreletter
\title{Characterization and Automatic Update of Deprecated Machine-Learning API Usages}

\author{Stefanus Agus Haryono, Ferdian Thung, David Lo, Julia Lawall, Lingxiao Jiang%
\IEEEcompsocitemizethanks{
\IEEEcompsocthanksitem S. A. Haryono, F. Thung, D. Lo, and L. Jiang are with School of Information Systems, Singapore Management University, Singapore.\protect\\E-mail:\{stefanusah,ferdianthung,davidlo,lxjiang\}@smu.edu.sg
\IEEEcompsocthanksitem J. Lawall is with Inria, France.\protect\\E-mail:Julia.Lawall@inria.fr}%
}

\markboth{Journal of \LaTeX\ Class Files,~Vol.~14, No.~8, August~2015}%
{Shell \MakeLowercase{\textit{et al.}}: Bare Demo of IEEEtran.cls for Computer Society Journals}

\IEEEtitleabstractindextext{%
\begin{abstract}
Due to the rise of AI applications, machine learning libraries have become far more accessible, with Python being the most common programming language to write them.
Machine learning libraries tend to be updated periodically, which may deprecate existing APIs, making it necessary for developers to update their usages.
However, updating usages of deprecated APIs is typically not a priority for developers, leading to widespread usages of deprecated APIs which expose library users to vulnerability issues. 
In this paper, we built a tool to automate these updates. We first conducted an empirical study to seek a better understanding on how updates of deprecated machine-learning API usages in Python can be done. The study involved a dataset of \numcasestudy{} deprecated APIs from {\tt Scikit-Learn, TensorFlow}, and {\tt PyTorch}.
We found dimensions of deprecated API migration related to its update operation (i.e., the required operation to perform the migration), API mapping (i.e., the number of deprecated and its corresponding updated APIs), and context dependency (i.e., whether we need to consider surrounding contexts when performing the migration). 
Guided by the findings on our empirical study, we created \toolname{}, a tool to automate the update of Python deprecated API usage that automatically infers the API migration transformation through comparison of the deprecated and updated API signatures. These transformations are expressed in a Domain Specific Language (DSL).
We evaluated \toolname{} using test dataset containing 258 files with 514 API usages that we collected from public GitHub repositories.
In this evaluation, \toolname{} achieves a precision of 86.19\%. 
We further improve the precision of \toolname{} by adding a feature that allows it to accept additional user input to specify the transformation constraints in the DSL for context-dependent API migration.
Using this addition, \toolname{} achieves a precision of 93.58\%, which is a 7.39\% precision improvement from our original solution.

\end{abstract}

}

\maketitle

\IEEEdisplaynontitleabstractindextext

\IEEEpeerreviewmaketitle

\IEEEraisesectionheading{\section{Introduction}}
The popularity of machine learning has surged in recent years, among both developers and researchers.
Currently, the most popular programming language for machine learning is Python, due to the vast amount of Python libraries that support and provide machine learning capability.
Furthermore, these libraries offer easy to use APIs, which allow developers to utilize machine learning without knowing the low-level implementation details.

Libraries are typically updated from time to time to add new features, fix bugs, or improve performance. With each new update, changes in the library's APIs are inevitable, which may include API deprecations. Usages of deprecated APIs must be replaced with their corresponding updated APIs to ensure that the code works with the future version of the library. Such deprecation also occurs in Python machine learning libraries.
Using the latest version of machine learning library is important for the performance and security of the program. Han et al.~\cite{han_dependency_deep_neural_network} stated that to take full advantage of deep learning libraries, users must always keep up to date with the latest library versions.
A study by Wang et al.~\cite{Lili_deprecated_API} supports a similar argument, stating that using an older version of deep learning library may expose library users to security issues and worse performance.

However, updating usages of deprecated APIs can be cumbersome and time-consuming.
This affect some developers to prefer using the older versions of a library, as demonstrated by several studies~\cite{Lili_deprecated_API, kula_do_developers_update, han_dependency_deep_neural_network}.
Hence, providing a better alternative to manually updating the usages of Python deprecated machine learning library APIs can benefit many developers.
Although deprecation of machine learning library APIs is common, each deprecation affects the API usages differently. Some deprecations can be fixed easily while others may require a deep understanding of the workflow or functionality of the API. 
No studies have attempted to get a better understanding on the deprecated API migration pattern for Python machine-learning libraries.

Prior to this study, several attempts have been made to automate the update of deprecated API usages in other programming languages. 
A4~\cite{lamothe_a4} is an approach for automatic Android API migration by learning API migration patterns from code examples.
AppEvolve\cite{fazzini2019automated}, provides automated API usage update for Android by using before- and after- update code example to create an update patch. Extending AppEvolve,
CocciEvolve\cite{coccievolve} provides an automated API usage update for Android using only a single after-update example, providing an easily readable transformation expressed in the form of semantic patch language (SmPL).
NEAT~\cite{thung_towardsgenerating} generates transformation rules for deprecated Android API usage replacements without any usage example. NEAT constructs signature graph to do the type conversions required by the API replacement.
No previous work has targeted Python API deprecation.

In this study, we followed an "as simple as possible, but no simpler" approach -- as advocated in many recent studies~\cite{fu_easy_over_hard, liu_neural_machine_translation, majumder_faster_than_deep_learning} -- by first doing an empirical study to ascertain the level of innovation needed and then designing a suitably complex solution to address the problem at hand. 
Through this empirical study, we aim to discover deprecated API migration dimensions in popular Python machine learning libraries. Within each dimension, we define the categories of deprecated API migrations that exist in Python machine learning libraries.

For our case study, we picked a subset of the available Python machine libraries based on the number of GitHub repositories that depend on the library (obtained through GitHub dependency graph\footnote{https://docs.github.com/en/github/visualizing-repository-data-with-graphs/about-the-dependency-graph}). 
According to this statistic, as of August 2020, more than 120,000 repositories utilize {\tt Scikit-learn}, while more than 82,000 repositories use {\tt Tensorflow}. {\tt Keras} and {\tt Pytorch} each have at least 52,000 and 34,000 utilizing repositories respectively.
However, since {\tt Keras} is included within {\tt TensorFlow} as of January 2017,\footnote{https://www.fast.ai/2017/01/03/keras/} we picked {\tt Scikit-Learn, TensorFlow}, and {\tt PyTorch} as our case study libraries.
We collected API deprecation messages from the documentation and changelogs of the last two years' (July 2018 - August 2020) major and minor releases of these libraries, where we found a total of \numcasestudy{} distinct APIs that were deprecated. We utilized manual thematic analysis and found three common dimensions in deprecated API migration, namely the chosen
update operation (based on the type of operation done in the API migration), the API mapping (based on the number of deprecated and updated APIs), and the need for context dependency (based on whether the API migration depends on information in the surrounding usage context). We also labelled each of the collected deprecated APIs based on the categories within each dimension to learn their distribution.


Based on the findings in the empirical study, we propose a tool to automate the update of Python deprecated API usages called \toolname{}. \toolname{} takes as input the deprecated API signature, the updated API signature, and the file to be updated.
It then automatically infers the required transformation by comparing the deprecated and updated APIs' signatures. The inferred transformations are expressed in the form of a Domain Specific Language (DSL) that we built based on the results of our empirical study. 

A comparison of features between \toolname{} and other similar refactoring and automatic API usage update tools \cite{fazzini2019automated, coccievolve, thung_towardsgenerating, lamothe_a4} is shown in Table~\ref{table:tools_comparison}. Compared to the tools from previous studies, \toolname{} is the only tool that supports the Python language along with its unique features.
Python has several unique features, including parameter default values, two types of parameters (positional parameter and keyword parameter), the usage of whitespaces and indentation as separator, and dynamic typing.
A keyword parameter is a type of parameter that can only be addressed through its name, which is in contrast with positional parameter that can be addressed through both its position and name.
The usage of whitespaces and indentation as separator makes parsing Python code different from other languages which typically depends on a specific symbol (e.g. Java and C which use semicolons and brackets).
Among the previous tools, the closest work to ours is NEAT\cite{thung_towardsgenerating}. However, the NEAT algorithm does not work well for Python. NEAT requires the parameter types of all API within a library to perform the automatic update. The need of parameter types for all API within the library means that we also need to label the type for APIs that are not deprecated. Due to the dynamic nature of Python, we would need to do this parameter types labelling manually, which renders NEAT no longer fully automatic.

\begin{table}[t]
\caption{Comparison of features between various automated API usage update tools}
\begin{center}
\begin{tabular}{|p{2.5cm}|K|J|L|}
\hline
                               & MLCatchUp                  & A4, AppEvolve, CocciEvolve               & NEAT                     \\ \hline
Automated API usage update     & \checkmark & \checkmark & \checkmark \\ \hline
Requires API migration example &                           & \checkmark &                           \\ \hline
Default parameter support      & \checkmark &                           &                           \\ \hline
Keyword parameters support     & \checkmark &                           &                           \\ \hline
Python language support        & \checkmark &                           &                           \\ \hline
\end{tabular}
\end{center}
\label{table:tools_comparison}
\end{table}


We evaluated \toolname{} using a dataset that we collected from public GitHub repositories using our prototype Python API usage GitHub search tool.
Our evaluation dataset is comprised of 258 files containing 514 deprecated API usages from 66 different Python machine learning library APIs. The update results created by \toolname{} are manually labelled by Python programmers with at least 4 years of experience who are not co-authors of this study. Based on the evaluation result, \toolname{} achieves a precision of 86.19\%.
We also conducted a qualitative study on the failed update results, where we found one of the main faults to be \toolname{}'s inability to infer transformation constraints for context-dependent deprecated API migrations.
To mitigate this problem, we add a feature into \toolname{} to accept additional user input in the form of a transformation constraint. Using this feature, we conducted a follow-up experiment where \toolname{} achieved 93.58\% precision.

\toolname{} is beneficial for library users, library developers, and future researchers who aim to ease the process of updating deprecated API usages. For library users, \toolname{} can be used to assist them in updating the usages of deprecated API within their codebase, significantly reducing the time required to update into a newer version of a library. 
Library developers can create a script that makes use of \toolname{} to update deprecated API usages within code. This script can then be distributed alongside the update of the library, making it easier for users to update the library.
For future researchers, \toolname{} can be a foundation for studies and tools on automated deprecated API usage update, especially for the Python programming language.

In summary, the main contributions of this work are as follows:
\begin{itemize}
    \item We collect a list of Python deprecated APIs from the releases over the last two years of {\tt Scikit-Learn, TensorFlow}, and {\tt PyTorch} by manually reading their major and minor releases changelogs.
    \item We define the dimensions of Python deprecated API usage migrations and compute the distributions for each dimension's categories to determine which type of API usage migration should be prioritized in the creation of an update tool.
    \item We create \toolname{}, an automated Python API usage update tool that can infer the required API transformation automatically. To the best of our knowledge, we are the first to create such a tool for Python.
    \item We evaluate \toolname{} using our manually collected dataset of 514 deprecated API usages collected from public Github repositories, where \toolname{} achieves 86\% precision.
\end{itemize}

The rest of this paper is organized as follows. Section~\ref{section:empirical_study} discusses our empirical study of Python deprecated API usage migrations and our findings from this study. Section~\ref{section:automated_update_tool} talks about the structure and features of our Python API usage automated update tool, \toolname{}. Section~\ref{section:tool_evaluation} describes our evaluation of \toolname{}, including the dataset used, the evaluation results, and our findings. Section~\ref{section:threats_to_validity} talks about the threats to validity of our work. Section~\ref{section:related_work} discusses related works on API deprecation and program transformation. Finally, Section~\ref{section:conclusion_future_work} summarizes and concludes our work and talks about the possibilities for future work.

\section{Empirical Study}\label{section:empirical_study}
Through this empirical study, we seek to get a better understanding of the categories of deprecated API migrations when updating usages of deprecated machine-learning APIs in Python.
We also measure the distribution of these categories to better comprehend which categories are the most relevant to Python machine-learning libraries.
Using these findings and analysis, our aim is to create a tool that can automate the process of updating the usage of deprecated Python machine learning APIs. 



The process of our empirical study is as follows. First, we define the research questions that we attempt to answer. 
Second, we select the libraries and APIs that we use as our case study and define our approach in collecting those APIs.
Third, we analyze the categories of deprecated API migration from our case study to answer our first research question. 
Finally, we measure the distribution of the deprecated API migration categories within our case study APIs to answer the second research question.

\subsection{Research Questions}
In this empirical study, we analyze the API migration dimensions of Python machine learning libraries and their prevalence in these libraries.
We ask the following research questions: 
\begin{itemize}[nosep,leftmargin=*]
    \item \textbf{RQ1} \textit{What are the dimensions of the API migrations for Python machine-learning deprecated API usages?}
    
    Currently, the API migration dimensions in deprecated Python APIs are unknown.
    Understanding these dimensions and their categories would help us in designing an automatic deprecated API migration tool. 
    
    \item \textbf{RQ2} \textit{What is the distribution of deprecated API migrations in Python machine-learning libraries?}
    
    Understanding the frequency of each API migration dimension and its categories will help researchers to determine the most common API migrations. Having such an understanding is beneficial in creating tools to ease the process of updating deprecated APIs.
    Information on the frequency of each API migration dimension's categories will help researchers and API migration tool developers in determining what type of API migration should be prioritized in the creation of an update tool.

    
\end{itemize}

\subsection{Case Study Data Collection}\label{section:case_study_data}

One of the main problems in collecting API deprecations from Python machine-learning libraries is that there are no clear specifications or standards on how the deprecations are documented. Each of these machine learning libraries has a different way of documenting their API deprecations. Moreover, even within the same library, the documentation for API deprecation varies.
While Python provides a warning function that can be used to mark a deprecated API,\footnote{https://docs.python.org/3/library/warnings.html} it is not commonly used by API developers.

As a consequence of the varying deprecation documentation, there is no automatic approach capable of listing the deprecations of machine learning library APIs. Furthermore the documentation and changelogs of all of the libraries that we consider are in the form of unstructured text, making it even harder to automatically collect the API deprecations.
Thus, we collected the API deprecations by manually reading the changelog of each library. To ease our search, we focused on looking for text containing the word \hbox{"deprecat-"} and \hbox{"replace-"}, which indicates API deprecation or replacement.
For {\tt scikit-learn}, the changelog is available in their official documentation page.\footnote{https://scikit-learn.org/stable/whats_new/v0.23.html} For {\tt Pytorch}{\footnote{https://github.com/pytorch/pytorch/releases}} and {\tt TensorFlow},{\footnote{https://github.com/TensorFlow/TensorFlow/releases}} the changelog can be found in their GitHub release page.

We thoroughly read the documentation provided by each library to collect a list of API deprecations and their updated APIs. We limit our manual collection effort to the changelogs of major and minor releases in the last two years (July 2018 - August 2020). 
For {\tt scikit-learn}, we read the changelogs of versions 0.21.0, 0.22.0, and 0.23.0. For {\tt Pytorch}, our study includes the changelogs of versions 0.4.1, 1.0.0, 1.1.0, 1.2.0, 1.3.0, 1.4.0, 1.5.0, and 1.6.0. Lastly, we included the changelogs of {\tt TensorFlow} versions 1.9.0, 1.10.0, 1.11.0, 1.12.0, 1.13.0, 1.14.0, 1.15.0, 2.0.0, 2.1.0, 2.2.0, and, 2.3.0.

From the documentation, we found a total of \numcasestudy{} pairs of deprecated APIs and their updated APIs, distributed across the libraries as shown in Table~\ref{table:deprecation_statistic}.
{\tt TensorFlow} has the highest number of deprecated APIs due to a significant update from version 1.x to 2.x, which introduced a compatibility API module. This compatibility module provides an interface to make use of deprecated APIs from the older version of TensorFlow. The creator of {\tt TensorFlow} recommends to immediately update any use of the compatibility module API to the newer API in the 2.x versions.\footnote{https://www.tensorflow.org/guide/upgrade}


\begin{table}[t]
\caption{Number of Deprecated APIs in Machine-Learning Libraries}
\begin{center}
\begin{tabular}{ |l|c| }

\hline
 \textbf{Library} & \textbf{\# Deprecated APIs}
\\ \hline
Scikit-Learn & 22
\\ \hline
TensorFlow & 54
\\ \hline
PyTorch & 36
\\ \hline
Total & 112
\\ \hline
\end{tabular}
\end{center}
\label{table:deprecation_statistic}
\end{table}

\subsection{RQ1: Deprecated API Migration Dimensions }\label{section:research_question_1}


In this research question, we aim to find the deprecated API migration dimensions in Python machine-learning libraries. 
To answer this research question, we utilized thematic analysis, which is a qualitative data analysis method for systematically identifying, organizing, and offering insight into patterns or themes across a dataset\cite{braun_thematic_analysis}. 
We followed the six steps of thematic analysis using the \numcasestudy{} deprecated APIs that we collected. These steps are as follows:

\begin{enumerate}
    \item {\em Familiarization with the data}. Familiarization with the data is done by thoroughly reading and analyzing the deprecated API migration data that we collected. In our study, data refers to the deprecated and updated signatures of the \numcasestudy{} deprecated APIs that we collected.
    \item {\em Initial coding of the data}. Code refers to the patterns or actions of the API migrations that are done to update the usage of deprecated APIs.
    \item {\em Searching for themes}. In our study, the themes that we aim to collect are the dimensions of deprecated API migrations. In this step, we search for the common dimensions of API migration that appear in multiple different deprecated APIs.
    \item {\em Reviewing themes}. We review the dimensions of API migrations that we found to ensure that all of the found dimensions are relevant. Similar dimensions are merged, while irrelevant dimensions are discarded.
    \item {\em Defining and naming themes}. We define the dimensions of deprecated API migrations and provide an informative name for each dimension that reflects its themes and meaning.
    \item {\em Writing up the results}. Finally, we write the result of our analysis, which is compiled in the empirical study of this paper.
\end{enumerate}



We found three dimensions of deprecated API migration, which are related to the {\em update operation} required to perform the migration, the {\em API mapping} from the deprecated to the updated APIs, and the {\em context dependence} of the migration (see Section~\ref{section:rq2_distribution} for the distribution of these dimensions). 
The definitions of these dimensions and their categories are as follows:

\noindent \textbf{Update Operation.}
We observe common patterns in the updates required to migrate a deprecated API to its updated APIs by looking at the differences between the signatures of the two APIs. We found the following categories of update operations:

\begin{itemize}[nosep,leftmargin=*]
\item \textbf{Remove parameter:} This operation removes the deprecated API usage function parameter(s). 
An example of this update operation can be seen in Figure~\ref{fig:remove_param_example}. In this example, the API {\tt sklearn.cluster.KMeans} has the deprecated parameter {\tt n_jobs} in line 2. After the update, the parameter is removed, with the result shown in line 3.
        
\begin{figure}
\centering
\scriptsize{
\begin{lstlisting}[xleftmargin=5.0ex,numbers=left,language=diff,sensitive=true,columns=flexible,basicstyle=\ttfamily]
    from sklearn.cluster import KMeans
-   clusterer = KMeans(n_clusters = 3, n_jobs = 1)
+   clusterer = KMeans(n_clusters = 3)
\end{lstlisting}
	\caption{Remove parameter operation to migrate deprecated {\tt sklearn.cluster.KMeans} API usage
	from {\tt scikit-learn} version 0.23.0}\label{fig:remove_param_example}
}
\end{figure}

\item \textbf{Rename parameter:} This operation replaces one or more keyword parameter names with a new name. Similar to remove parameter deprecation, this update operation only affect API usages where the deprecated keyword parameter is explicitly declared.
        
\item \textbf{Convert positional parameter to keyword parameter:} This operation removes the positional parameter(s) and uses their value(s) to create new keyword parameter(s) within the API usage. 
An example of this update operation can be seen in Figure~\ref{fig:pos_to_key_example}. In this example, the {\tt torch.add} API's second positional parameter is deprecated and replaced with the keyword parameter {\tt alpha}. The deprecated API usage can be seen in line 4, while the updated code can be seen in line 5.

\begin{figure}
\centering
\scriptsize{
\begin{lstlisting}[xleftmargin=5.0ex,numbers=left,language=diff,sensitive=true,columns=flexible,basicstyle=\ttfamily]
    import torch
    t_0 = torch.randn((3, 3))
    t_1 = torch.ones_like(t_0)
-   t_add = torch.add(t_0, 10, t_1)
+   t_add = torch.add(t_0, t_1, alpha=10)
\end{lstlisting}
	\caption{Positional to keyword parameter operation to migrate deprecated {\tt torch.add} API usage from {\tt PyTorch} version 1.5.0}\label{fig:pos_to_key_example}
}
\end{figure}
    
\item \textbf{Rename a method:} This operation renames the function and/or the module of the API that is deprecated into a new updated name.
Figure~\ref{fig:rename_API_example} shows an example of this update operation for a {\tt torch.gels} API usage in line 3. The {\tt torch.gels} method name is deprecated and is replaced with the new name {\tt torch.lstsq}. 

\begin{figure}
\centering
\scriptsize{
\begin{lstlisting}[xleftmargin=5.0ex,numbers=left,language=diff,sensitive=true,columns=flexible,basicstyle=\ttfamily]
    import torch
    def solve(A, b, out=None, bias=True):
-       x, _ = torch.gels(b, A)
+       x, _ = torch.lstsq(b, A)
\end{lstlisting}
	\caption{Rename method operation to migrate deprecated {\tt torch.gels} API usage from {\tt PyTorch} version 1.3.0}\label{fig:rename_API_example}
}
\end{figure}



\item \textbf{Add a parameter:} This operation adds a positional/keyword parameter to an API invocation, as illustrated in Figure~\ref{fig:add_parameter_example}. In lines 3-5, we can see the deprecated {\tt torch.nn.functional.affine_grid} API usage, in which the keyword parameter {\tt align_corners} is not specified. To update this deprecated API usage, a new keyword parameter {\tt align_corners} needs to be added as can be seen in lines 6-7.

\begin{figure}[t]
\centering
\scriptsize{
\begin{lstlisting}[xleftmargin=5.0ex,numbers=left,language=diff,sensitive=true,columns=flexible,basicstyle=\ttfamily]
    def forward(self, input_image, affine_params):
        affine_params = affine_params.view(-1, 2, 3)
        affine_grid = torch.nn.functional.affine_grid(
            affine_params, (input_image.size(0), 
-           input_image.size(1), *self.img_size))
+           input_image.size(1), *self.img_size),
+           align_corners=True)
\end{lstlisting}
	\caption{Add parameter operation to migrate deprecated {\tt torch.nn.functional.affine_grid} API usage from {\tt PyTorch} version 1.3.0}\label{fig:add_parameter_example}
}
\end{figure}




\item \textbf{Change a parameter type:} This operation changes the type of a deprecated API usage parameter. 
The operation first checks whether the type of the current API argument matches the new type of the updated API. 
If the current argument type does not match the new type, the deprecated API argument needs to be changed to follow the new type specification.
In this operation, a previously valid type may no longer be usable in the parameter.  
The creation of new variables or values may be needed in this update operation.
Figure~\ref{fig:parameter_constraint_example} provides an example of this operation for {\tt sklearn.utils.estimator_checks.} {\tt check_estimator} deprecated API usage. In this example, the allowed parameter type in the updated API is changed to only the {\tt Estimator} type. To update this deprecated API usage, a type check is added for the function invocation argument (lines 5-8) to ensure that its value is suitable for the updated API. 

\begin{figure}[t]
\centering
\scriptsize{
\begin{lstlisting}[xleftmargin=5.0ex,numbers=left,language=diff,sensitive=true,columns=flexible,basicstyle=\ttfamily]
    import sklearn
    def check_estimator(TestEstimator):
-       sklearn.utils.estimator_checks.check_
-               estimator(SklearnWrapperClassifier)
+       if isinstance(TestEstimator, 
+               sklearn.base.BaseEstimator):
+           sklearn.utils.estimator_checks.check_
+                   estimator(SklearnWrapperRegressor)
\end{lstlisting}
	\caption{Migration for {\tt sklearn.utils.estimator_} {\tt checks.check_estimator} deprecated API usage from {\tt scikit-learn} version 0.23.0}\label{fig:parameter_constraint_example}
}
\end{figure}


\item \textbf{Add a constraint to a parameter value:} This operation adds a constraint to the value of the API parameter due to a change in the permitted value of the parameter.
This operation adds a value check for the API argument to ensure that the current value of the argument fits the newly permitted range of values of the API parameter. If the current argument does not fit, we need to modify its value accordingly.
An example of this update operation is shown in Figure~\ref{fig:value_constraint_example} for {\tt torch.normal} deprecated API usage. In this example, the value of the {\tt out} argument must have the same size as the {\tt mean} and {\tt std} arguments. To update the usage of this deprecated API, we add an if statement that checks the sizes of these objects (line 4-5)

\begin{figure}
\centering
\scriptsize{
\begin{lstlisting}[xleftmargin=5.0ex,numbers=left,language=diff,sensitive=true,columns=flexible,basicstyle=\ttfamily]
    import torch
    def normal(mean, std, output):
-       torch.normal(mean, std, out=output)
+       if mean.size()==output.size() and 
+               std.size()==output.size():
+           torch.normal(mean, std, out=output)

\end{lstlisting}
	\caption{Migration for {\tt torch.normal} deprecated API usage from {\tt torch} version 1.5.0}\label{fig:value_constraint_example}
}
\end{figure}

\item \textbf{Remove API:} This operation removes the deprecated API without replacing its usage with any updated API. This is typically used when the deprecated API is no longer needed or no longer has any effect when invoked. 
Figure~\ref{fig:remove_api_example} shows an example of this update operation for {\tt sklearn.linear_model.logistic_regression_} {\tt path}. To update this deprecated API usage, the deprecated API is removed (line 2).

\begin{figure}[t]
\centering
\scriptsize{
\begin{lstlisting}[xleftmargin=5.0ex,numbers=left,language=diff,sensitive=true,columns=flexible,basicstyle=\ttfamily]
    from sklearn.linear_model import 
        logistic_regression_path
-   classificador = logistic_regression_path(
-       X=previsores_teste, y=classe_treinamento)
\end{lstlisting}
	\caption{Migration for {\tt sklearn.linear_model.logistic_} {\tt regression_path} deprecated API usage from {\tt sklearn} version 0.21.0}\label{fig:remove_api_example}
}
\end{figure}


\end{itemize}

\hspace*{\fill}

\noindent \textbf{API Mapping.}
The type of mapping represents the ratio between the number of the deprecated and updated APIs involved in the migration. 
We found three types of mappings from our investigation:

\begin{itemize}[nosep,leftmargin=*]
    \item \textbf{1:1 API mapping:} It occurs when a deprecated API is modified or replaced by a single updated API.
    \item \textbf{1:N API mapping:} It occurs when a deprecated API is modified or replaced by at least two updated APIs.
    \item \textbf{1:0 API mapping:} It occurs when a deprecated API is removed without any suggested replacement.
\end{itemize}

\hspace*{\fill}

\noindent\textbf{Context Dependency.}
It indicates whether the deprecated API migration depends on the context of the deprecated API usage. This context refers to the value and type of the arguments in the deprecated API usage. We found two categories of context dependency.
\begin{itemize}[nosep,leftmargin=*]
\item \textbf{Context-dependent update}. It is a migration where the required update depends on the value of one or more arguments of the API invocation. 
An example of this case is found in the {\tt sklearn.model_selection.KFold} API, where the usage is only deprecated if the {\tt random_state} argument is not {\tt None}, in which case the {\tt shuffle} argument must be set to {\tt True}. Otherwise, there is no change to the API.
For example, in Figure~\ref{fig:context_dependent_example}, the update is done by adding an {\tt if} statement that checks the value of the contextual variable {\tt seed} used as the {\tt random_state} argument.

\item \textbf{Context-independent update}. This migration is not affected by the value of the API invocation arguments.
\end{itemize}

\begin{figure}
\centering
\scriptsize{
\begin{lstlisting}[xleftmargin=5.0ex,numbers=left,language=diff,sensitive=true,columns=flexible,basicstyle=\ttfamily]
    from sklearn.model_selection import KFold
    def KFold_Selection(seed):
+       if seed is not None:
            KFold(random_state = seed, shuffle=True)
+       else:
+           KFold(random_state = seed)
    
    \end{lstlisting}
    	\caption{Context-dependent change deprecated API migration example for {\tt sklearn.model_selection.KFold} from {\tt scikit-learn} version 0.22.0}\label{fig:context_dependent_example}
    }
    \end{figure}

\subsection{RQ2: Migration Dimensions Distribution}\label{section:rq2_distribution}
In the second research question, we are interested in the distribution of deprecated API migration categories within the dimensions defined in Section~\ref{section:research_question_1}: update operation, API mapping, and context dependency. For this purpose, we labelled each of the \numcasestudy{} APIs based on the three dimensions and their categories.

First we label the APIs based on the update operation. The result of this labelling can be seen in Table~\ref{table:update_operation_statistic}. 
The most common operations vary according to the library.
We found that most deprecated API migrations in {\tt scikit-learn} involve removing parameter.
{\tt PyTorch} and {\tt TensorFlow} deprecated API migrations mainly rename methods. 
{\tt TensorFlow} API migrations also often rename parameters. Among the update operations, change parameter type, add constraint to parameter value, and remove API are the least commonly used, amounting to 5 or fewer APIs for each update operation. 

\begin{table}
\centering
\caption{Distribution of the case study API based on the required update operation to perform the migration}
\label{table:update_operation_statistic}
\begin{tabular}{|p{3cm}|c|c|c|}
\hline
\textbf{}              & \multicolumn{1}{l|}{\textbf{Scikit-learn}} & \multicolumn{1}{l|}{\textbf{PyTorch}} & \multicolumn{1}{l|}{\textbf{TensorFlow}} \\ \hline
Remove Param           & 12                                         & 1                                     & 0                                        \\ \hline
Rename Param           & 0                                          & 1                                     & 25                                       \\ \hline
PosToKey Param         & 0                                          & 8                                     & 0                                        \\ \hline
Rename Method          & 4                                          & 18                                    & 17                                       \\ \hline
Add Parameter          & 2                                          & 1                                     & 9                                        \\ \hline
Change Param Type      & 2                                          & 3                                     & 0                                        \\ \hline
Add Param Value Constraint & 1                                          & 4                                     & 0                                        \\ \hline
Remove API            & 1                                          & 0                                     & 3                                        \\ \hline
\end{tabular}
\end{table}

We also label each API based on its API mapping, as shown in Table~\ref{table:update_replacement_statistic}. Almost all deprecated API migrations follow the 1:1 transformation, suggesting that API developers prefer a straightforward, less complex API update. Out of the \numcasestudy{} APIs, only 1 API follows the 1:N update category. The 1:0 category is found in 4 APIs.

\begin{table}
\centering

\caption{Distribution of the case study API based on its API mapping}
\begin{tabular}{|c|c|c|c|}
\hline
\textbf{} & \textbf{Scikit-learn} & \textbf{PyTorch} & \textbf{TensorFlow} \\ \hline
1:1       & 21                    & 36               & 50                  \\ \hline
1:N       & 0                     & 0                & 1                   \\ \hline
1:0       & 1                     & 0                & 3                   \\ \hline
\end{tabular}
\label{table:update_replacement_statistic}
\end{table}

Then, we label each API based on its update context.
The results of this labelling can be seen in Table~\ref{table:context_based_statistic}. The majority of the API updates are context-independent. Out of \numcasestudy{} APIs, only 9 API updates are context-dependent.

\begin{table}[ht]
\centering

\caption{Distribution of the case study API based on the contextual dependency of the API update}
\begin{tabular}{|c|c|c|c|}
\hline
\textbf{}           & \textbf{Scikit-learn} & \textbf{PyTorch} & \textbf{TensorFlow} \\ \hline
Context-independent & 19                    & 30               & 54                  \\ \hline
Context-dependent   & 3                     & 6                & 0                   \\ \hline
\end{tabular}
\label{table:context_based_statistic}
\end{table}

From these distributions within the three dimensions we found that simpler updates are preferred among different Python machine learning libraries. Rename method, rename parameter, add parameter, and remove parameter are the most commonly used, each with at least 10 APIs using the approach. 107 out of \numcasestudy{} APIs follow the 1:1 update mapping, suggesting a tendency towards a less complex updates. Finally, majority of API updates are context-independent, which is the case for 103 out of \numcasestudy{} APIs.
We leverage these findings to guide our priorities in creating automated update tool for Python machine learning deprecated API usage.

\begin{figure*}[t]
	\centering
	\includegraphics[width=0.8\linewidth]{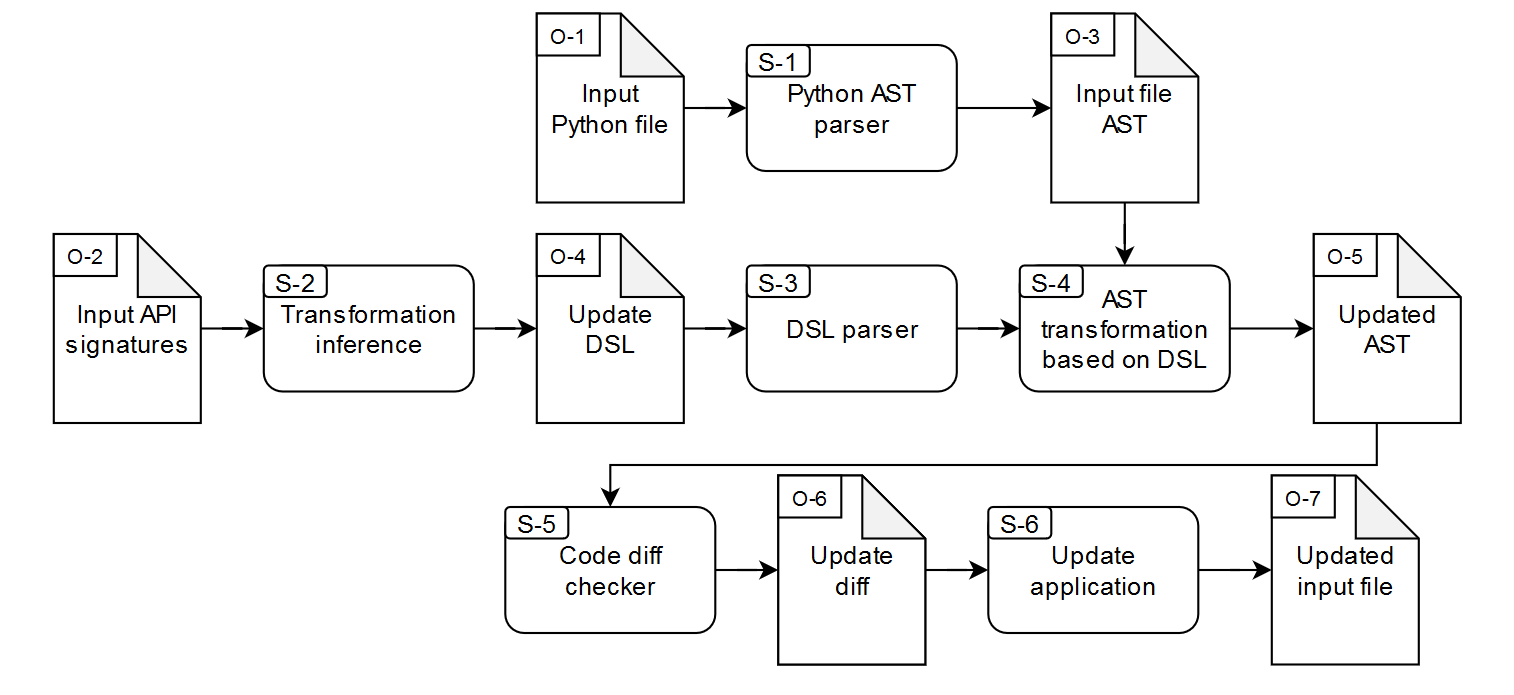}
	\caption{\toolname{} Architecture}
	\label{fig:machine_learning_transformation}
\end{figure*}

\section{Automated Update for Python Deprecated API Usages}\label{section:automated_update_tool}


Based on the findings of our empirical study and the fact that Python deprecated APIs are still being used by developers, we are interested in creating an automated approach to update the usage of Python deprecated APIs.
We incorporate the findings of our empirical study in creating this update tool, mainly to direct 
the types of transformations that the tool should support. 



We create \toolname{}, a tool that provide automated update for Python deprecated API usage. \toolname{} infers the required transformations for the update by comparing the difference between the deprecated API signature and the updated API signature. Using these API signatures and a code containing the usage of deprecated API as an input, \toolname{} will automatically provide a new version of the code where all usages of the specified deprecated APIs are replaced with the corresponding updated API. During the transformation process, \toolname{} utilizes a domain specific language to make its transformation more readable for the users.
In the following subsections, we discuss in detail on the architecture and features of \toolname{}.

\subsection{\toolname{} Architecture}

The architecture and pipeline of \toolname{} are given in Figure~\ref{fig:machine_learning_transformation}. There are two main inputs to \toolname{}, namely the input Python file (O-1), and the input API signatures (O-2). The input Python file is the file that is going to be updated, while the input API signatures are the deprecated API signature and the updated API signature. The input Python file is transformed into its AST using the abstract syntax tree module provided by Python\footnote{https://docs.python.org/3/library/ast.html} (S-1), creating the input file AST (O-3). 
From this AST, \toolname{} detects and locates the usage of the deprecated API based on the deprecated API signature. Our approach in this deprecated API detection is adapted from the work of Wang et al.~\cite{Lili_deprecated_API}, which is done through the normalization of all API usages into their fully qualified name, and comparisons of these names to the deprecated API signature.


Input API signatures are used in the transformation inference process (S-2) where the necessary transformations for the API migration are automatically inferred.
The necessary transformations are in the form of DSL commands (O-4), containing a series of atomic operations required for the update (e.g. rename method, rename parameter, etc.) which are to be performed sequentially.
The DSL is then parsed by the DSL parser (S-3) into a list of operations that can be executed by \toolname{}. These operations are applied to the input file AST (S-4), producing the updated AST (O-5).

The updated AST is then compared with the input file AST by using the code diff checker (S-5), which lists all the code differences between the two ASTs and outputs the update diff (O-6).
Finally, based on the code differences in the update diff (O-6), the update is applied to the input Python file (S-6) by making only the necessary changes to the API usages without any modification to the original code comments and spacing. This results in the updated input file (O-7).

In the following sections, we describe in detail the main components of the \toolname{} automated update process. 
Section~\ref{section:DSL_and_operations} explains the transformation operations provided by \toolname{} and their DSL equivalents. This section also describes the grammar of the DSL that is used by \toolname{}.
Section~\ref{section:transformation_inference} discusses how our approach creates the automatic transformation based on the deprecated and updated API signatures.

\setlength{\grammarparsep}{5pt plus 1pt minus 1pt} 
\setlength{\grammarindent}{18em} 

\begin{figure*}
\centering
\begin{grammar}
<transformations> ::= { <transformation_op> [ if <transformation_constraint> ]; }+

<transformation> ::= <rename_method> \alt <rename_parameter> \alt <remove_parameter> \alt <positional_to_keyword_parameter> \alt <add_parameter>

<rename\textunderscore method> ::= rename\textunderscore method <old_method> to <new_method>

<rename\textunderscore parameter> ::= rename\textunderscore parameter <old_parameter_name> to <new_parameter_name> for <old_method>

<remove\textunderscore parameter> ::= remove\textunderscore parameter <old_parameter_name> for <old_method>

<positional\textunderscore to\textunderscore keyword\textunderscore parameter> ::= positional\textunderscore to\textunderscore keyword position <parameter_position> keyword <new_parameter_name> for <old_method>

<add\textunderscore parameter> ::= add\textunderscore parameter <new_parameter_name> with\textunderscore value <expression> for <old_method>

<remove\textunderscore api> ::= remove\textunderscore api <old_method>

<transformation\textunderscore constraint> ::= <transformation_constraint> and <transformation_constraint> \alt <transformation_constraint> or <transformation_constraint> \alt [not] <old_parameter> <constraint_definition>

<constraint\textunderscore definition> ::= <type_constraint> \alt <value_constraint>

<type\textunderscore constraint> ::= has type <type>

<value\textunderscore constraint> ::= has value <expression>

<old\textunderscore method> $\in$ a set of deprecated API signature

<new\textunderscore method> $\in$ a set of updated API signature

<old\textunderscore parameter\textunderscore name> $\in$ a set of parameter name

<new\textunderscore parameter\textunderscore name> $\in$ a set of parameter name

<parameter\textunderscore position> $\in$ a set of positive number

<type> $\in$ a tfpdef\footnotemark

<expression> $\in$ a expr\textunderscore stmt\footnote[9]{https://docs.python.org/3.6/reference/grammar.html}

\end{grammar}
\caption{Grammar of the \toolname{} domain specific language}\label{fig:DSL_Grammar}
\end{figure*}

\subsection{Transformation Operations and DSL}\label{section:DSL_and_operations}

In providing the required update to mitigate the usage of a deprecated API, \toolname{} makes use of several basic transformation operations. We choose the required basic transformation operations based on the result of our empirical study, where we found that the varieties of deprecated API migration can be summarized into 8 different update operations, as described in Section~\ref{section:research_question_1}.

We leverage these findings as the building block of the automatic update functionality provided by \toolname{}.
Each operation within \toolname{} is paired with a corresponding construct in the domain specific language to describe the transformation.
The required transformation operations for each API migration are inferred automatically, as we will discuss in Section~\ref{section:transformation_inference}. 
This transformation inference process will produce a DSL script that specifies all the required transformations between a pair of deprecated and updated API.
The grammar of the DSL in EBNF (Extended Backus–Naur Form) is shown in Figure~\ref{fig:DSL_Grammar}.

Each transformation listed in the DSL grammar corresponds to a basic transformation operation provided by \toolname{}. The mapping between each basic transformation operation and its DSL grammar is shown in Table~\ref{table:operation_grammar_pairing}.

\begin{table}[t]
\centering
\caption{Mapping between each transformation operation and its DSL grammar}
\label{table:operation_grammar_pairing}
\begin{tabular}{|N|M|}
\hline
\textbf{Transformation   Operation}        & \textbf{DSL Grammar}                                        \\ \hline
Rename Method                             & \textless{}rename\textunderscore method\textgreater{}                     \\ \hline
Rename   Parameter                        & \textless{}rename\textunderscore parameter\textgreater{}                  \\ \hline
Remove   Parameter                        & \textless{}remove\textunderscore parameter\textgreater{}                  \\ \hline
Convert   Positional to Keyword Parameter & \textless{}positional\textunderscore to\textunderscore keyword\textunderscore parameter\textgreater{} \\ \hline
Add Parameter                             & \textless{}add\textunderscore parameter\textgreater{}                     \\ \hline
Change   Parameter Type                   & \textless{}type\textunderscore constraint\textgreater{}                   \\ \hline
Add Constraint   to Parameter Value       & \textless{}value\textunderscore constraint\textgreater{}                  \\ \hline
Remove API                                & \textless{}remove\textunderscore api\textgreater{}                        \\ \hline
\end{tabular}
\end{table}

\subsection{Transformation Inference}\label{section:transformation_inference}
\toolname{} automatically infers the required change to update the usage of the deprecated API using the deprecated and updated API signatures as its inputs. Based on these signatures, \toolname{} produces DSL commands containing the steps required to update code containing deprecated API usages.

\begin{figure}[t]
	\centering
	\includegraphics[width=0.98\linewidth]{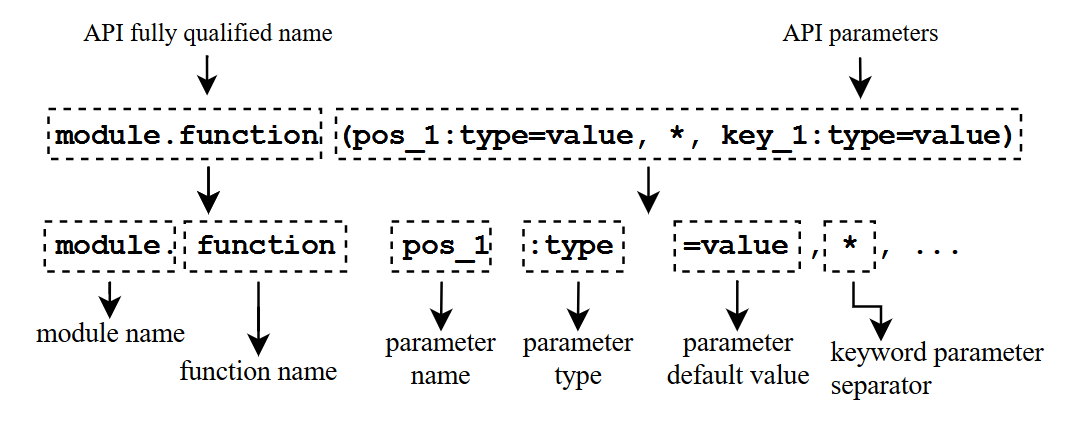}
	\caption{Syntax of \toolname{} API signature}
	\label{fig:api_signature_syntax}
\end{figure}

We define an API signature as the fully qualified name of the API and the list of parameters available for the API, where each parameter consists of its name, type, and an optional default value. A fully qualified API name consists of the API module names and the API function name.
This information is sufficient as Python does not provide method overloading.
The API parameter definition is declared in the form of a comma separated list of positional and keyword parameters, where each parameter has a specific type and may include a default value. To separate positional and keyword parameters, the symbol {\tt *} is used.
The syntax that is accepted by \toolname{} for API signatures is shown in Figure~\ref{fig:api_signature_syntax}.

\begin{algorithm}
\SetAlgoLined
\SetKwInOut{Input}{Input}
\SetKwInOut{Output}{Output}
\Input{$D \in$ deprecated API signature\newline $U \in$ updated API signature\;}
\Output{Update transformation DSL commands $Cmds$}
$Cmds \leftarrow null$\\
\eIf{$U$ $\equiv$ $null$}
{$Cmds.add(remove\textunderscore API(U_{name}))$}
{
{
\If{$D{name}$ $\neq$ $U_{name}$}
{$Cmds.add(rename\textunderscore method(D_{name}, U_{name}))$}
$D$, $U$ $\leftarrow$ $eliminate\textunderscore identical\textunderscore param(D,U)$\\

\ForEach
{Parameter P $\in$ $D_{param}$}
{
\ForEach{Parameter Q $\in$ $U_{param}$}
{

\While{$\exists$ ($P_{type}$ $\equiv$ $Q_{type}$ $\wedge$ $P_{name}$ $\neq$ $Q_{name}$)}
{
$Cmds.add(rename\textunderscore parameter(P, Q))$\\
$D$, $U$ $\leftarrow$ $remove\textunderscore used\textunderscore param(P,Q)$
}

\While{$\exists$ ($P_{type}$ $\equiv$ $Q_{type}$ $\wedge$ $P$ $is$ $a$ positional param $\wedge$ $Q$ $is$ $a$ keyword param)}
{
$Cmds.add($\\
    $positional\textunderscore to\textunderscore keyword\textunderscore parameter($\\
    $P,Q))$\\
$D$, $U$ $\leftarrow$ $remove\textunderscore used\textunderscore param(P,Q)$
}

\While{$\exists$ (P $\in$ $D_{param}$ $\wedge$ P $\notin$ $U_{param}$)}
{$Cmds.add(remove\textunderscore parameter(P))$\\}
\While{$\exists$ ($Q$ $\in$ $U_{param}$)}
{$Cmds.add(add\textunderscore parameter(Q))$}
}
}
}

}
\Return $Cmds$

 \caption{\toolname{} automatic inference}
 \label{algo:inference_pseudocode}
\end{algorithm}

The difference between the deprecated API signature and the updated API signature is used to automatically infer the required transformation for the API usage update. The pseudocode of this transformation inference is shown in Algorithm~\ref{algo:inference_pseudocode}.
The transformation inference takes as inputs the deprecated API signature and the updated API signature. Using these inputs, several steps are done to produce the DSL commands to update the deprecated API usage.
First, if the updated API signature is an empty string, we add a {\tt remove_api} operation (lines 2-3).
If the updated API signature is not empty, we compare the fully qualified name (line 5) of the deprecated and updated API signature, and add a {\tt rename_method} operation if the names differ (line 6).
Then, we eliminate identical API parameters from both API signatures (line 8), as the same parameters between the two APIs will not affect the API update. Two API parameters are considered identical if they have the same keyword or position, and the same type.

\footnotetext[9]{https://docs.python.org/3.6/reference/grammar.html}

After removing identical parameters, we compare the remaining API parameters from the two API signatures. Two possible actions are taken during this comparison. 
First, if two API parameters have the same type but different keyword names, we add a {\tt rename_parameter} operation into the DSL (lines 11-14).
Second, if two API parameters have the same type, but there is a positional parameter in the deprecated API signature and a keyword parameter in the updated API signature, we add a {\tt positional_to_keyword_parameter} operation (lines 15-18).
Then, we check whether there are any remaining API parameters in the deprecated and the updated API signatures.
Each of the remaining API parameters in the deprecated API signature is added as a {\tt remove_parameter} operation into the DSL commands (lines 19-21).
Finally, for each remaining API parameter in the updated API signature, we add an {\tt add_parameter} operation.

\begin{figure}[t]
	\includegraphics[width=0.90\linewidth]{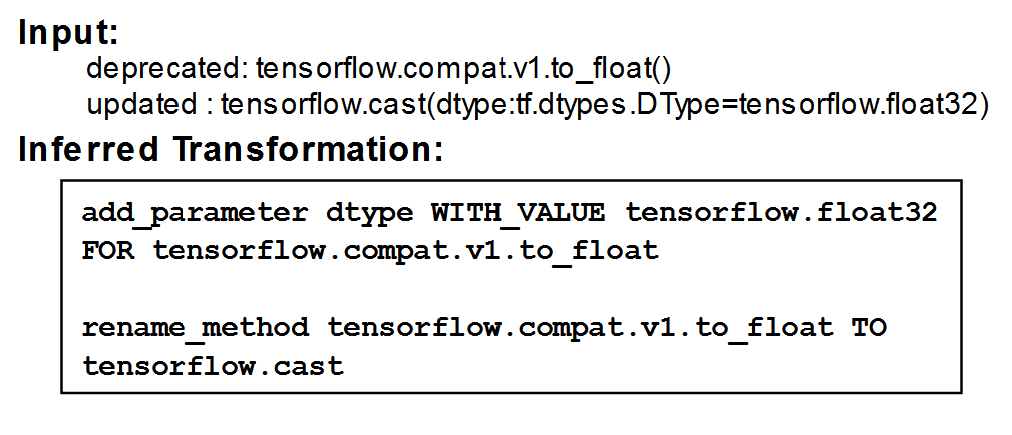}
	\caption{DSL commands produced to convert {\tt tensorFlow.compat.v1.to_float()} usages to use {\tt tensorflow.cast(dtype=tensorflow.float32)}}
	\label{fig:dsl_commands}
\end{figure}

As an illustration, consider the {\tt TensorFlow.compat.} {\tt v1.to_float()} deprecated API. This API usage needs to be updated to use the {\tt TensorFlow.} {\tt cast(dtype=TensorFlow.float32)} API. Using these API signatures, \toolname{} performs the transformation inference using the described steps. 
First, \toolname{} finds a difference between the fully qualified name of the deprecated API and the fully qualified name of the updated API, and thus a {\tt rename_method} operation is added.
Then, \toolname{} finds that there are no identical parameters between the two API signatures, leaving only a single API parameter in the updated API signature. This parameter is added as an {\tt add_parameter} operation.
The value of the argument added in the {\tt add_parameter} operation is obtained from the default parameter value listed in the API signature.
The produced DSL commands are shown in Figure~\ref{fig:dsl_commands}.

These DSL commands can then be applied to files containing the deprecated {\tt TensorFlow.compat.v1.to_float()} API usage. First, the file needs to be parsed into its AST. Then, each DSL command is parsed and translated into the corresponding transformation that is applied to the AST. After all the transformations are applied, the AST is changed back into the code form and the code differences between the updated code and the deprecated code are presented to the users. 
Based on these code differences, \toolname{} applies the transformation to the original code, making only the necessary changes towards the deprecated API usages, retaining the comments and spacing of the original code.
An example result of the update is shown in the code {\em diff} in Figure~\ref{fig:update_example}.

\begin{figure}
\centering
\scriptsize{
\begin{lstlisting}[xleftmargin=5.0ex,numbers=left,language=diff,sensitive=true,columns=flexible,basicstyle=\ttfamily]
import TensorFlow.compat.v1 as tf
+ from TensorFlow import cast
def parse_example(d):
    img = tf.decode_raw(d['image_raw'], tf.uint8)
-   d['image'] = tf.to_float(img)
+   d['image'] = cast(img, dtype=TensorFlow.float32)
    return d
\end{lstlisting}
	\caption{Migration result for {\tt TensorFlow.compat.v1.} {\tt to_float()} deprecated API usage from {\tt TensorFlow} version 2.0.0}\label{fig:update_example}
}
\end{figure}

\begin{figure}[t]
	\includegraphics[width=0.90\linewidth]{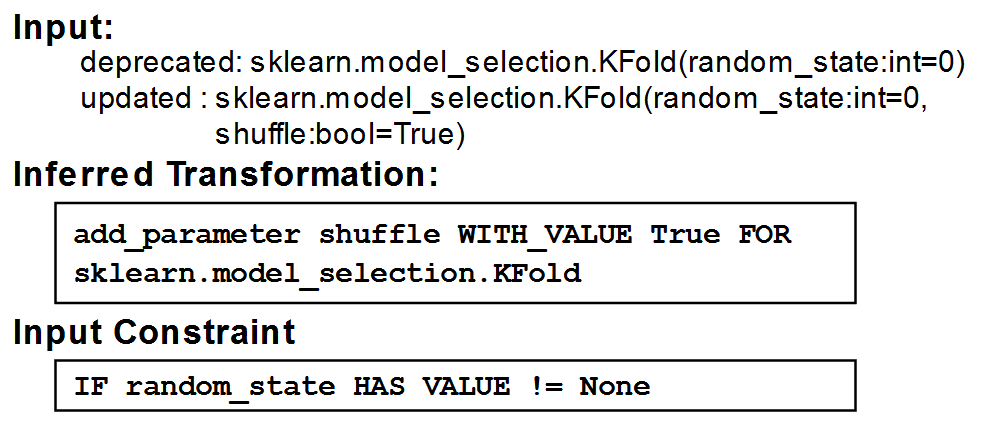}
	\caption{DSL commands with input constraint for migrating {\tt sklearn.model_selection.KFold} deprecated API usage from {\tt scikit-learn} version 0.23.0}
	\label{fig:constraint_DSL}
\end{figure}

While \toolname{} can infer the required transformation operations, \toolname{} is unable to infer the required information regarding any added constraints for the transformation itself.
Indeed the API signatures do not provide such information. 
To mitigate this problem, we allow \toolname{} to be run with an optional argument in which the user can specify the transformation constraint as an input. This input constraint must be written using the \toolname{} domain specific language. An example of the inferred DSL script and the provided input constraint for the {\tt sklearn.model_selection.KFold} API update is shown in Figure~\ref{fig:constraint_DSL}.

In Figure~\ref{fig:constraint_DSL}, we see that from the API signatures, \toolname{} is able to infer that {\tt sklearn.model_selection.KFold} deprecated API usage needs to be updated with the addition of {\tt shuffle:bool=True} parameter. However, \toolname{} is unable to obtain the transformation constraint information from the API signatures, providing an incomplete inferred series of transformation commands. In actuality, the add parameter update operation should only be done if the value of {\tt random_state} parameter is not None. This constraint is expressed through a user input constraint.


    
    

\section{\toolname{} Evaluation}\label{section:tool_evaluation}

To evaluate \toolname{}, we conducted an experiment to analyze the update results.
In this section, we explain the setting, data, and results of this evaluation.
We also conduct a qualitative study on the kinds of deprecated API usages that cannot be updated successfully by \toolname{} and discuss the possible approaches to mitigate them for future work.

\subsection{Dataset Collection}\label{section:dataset_collection}
Based on our empirical study, we use deprecated APIs from {\tt Scikit-learn, TensorFlow}, and {\tt Keras} libraries for our evaluation. 
We consider the \numcasestudy{} identified deprecated APIs from the three libraries.
In order to provide a realistic evaluation, we use real API usages for our evaluation. With this in mind, we directed our API usage search towards public GitHub repositories, which contain code from developers across the globe.

At first, we tried to use the GitHub code search API\footnote{https://developer.github.com/v3/search/} by using the deprecated API signature as the text-based search query. However, a lot of irrelevant results were returned, as the search cannot differentiate between API usages and other code elements. 
Accordingly, we created a prototype GitHub search engine that is able to accurately find code containing usage of Python APIs. 

We adapted the work of Wang et al.~\cite{Lili_deprecated_API} to build this search tool.
Our search tool makes use of the GitHub search API, allowing up to 1,000 repositories to be retrieved for each search query.
For the deprecated API usage detection, we utilized API normalization which rewrites all API usages in the code with their fully qualified name. Each fully qualified name is matched with the deprecated API signature to detect usages of deprecated APIs in the code. We also added multiprocessing capability to our search tool, reducing the execution time. 
Using the deprecated API signatures as the input, we collected our evaluation data.



\subsection{Evaluation Dataset}\label{section:evaluation_dataset}
Using our search tool, we collected code containing deprecated API usages from GitHub public repositories for the \numcasestudy{} case study APIs. Out of the \numcasestudy{} APIs, we found public code usages for only 66 APIs. 
For each of these 66 APIs, we collected at most five files containing deprecated API usage for our evaluation dataset.
For deprecated APIs used by more than five files, we randomly selected the five files. Meanwhile, for deprecated APIs used by only five or fewer files, we used all of the files for our evaluation.
In total, we collected 258 files containing 514 API usages from 66 different APIs.
45 files contain {\tt Scikit-learn} API usages, 89 files contain {\tt Tensorflow} API usages, and 124 files contain {\tt PyTorch} API usages.


\subsection{Experiment Setting}\label{section:experiment_setting}

We conducted an experiment to evaluate the ability of \toolname{} to update the usages of deprecated APIs.
We aim to measure how precise are the updates produced by \toolname{} through this experiment.
With this goal in mind, we calculated the precision of the update result. 
We define precision as the percentage of the correct updates performed during the test. 
The formulas for precision is shown below:

\begin{align*}
    Precision=\frac{TP}{TP+FP} \label{formula:Precision}
\end{align*}

Instances of deprecated API usages are automatically detected and updated using \toolname{} based on the input API signatures and the inferred transformations. For each updated API usage, there are three possible labels:
\begin{enumerate}[nosep,leftmargin=*]
    \item True Positive (TP): deprecated API usage is updated correctly
    \item False Positive (FP): deprecated API usage is updated incorrectly, i.e. the updated API usage is incorrect
\end{enumerate}
To label the update result, we employed the help of three Python programmers with at least four years of experience.
These programmers are non-authors, and were given instruction and training before they conducted the labelling process. 
Each instance of the deprecated API usages is labelled by two different programmers.
If any labelling disagreement arises between the two programmers, we conducted a discussion until a consensus is achieved.
Throughout the labelling process, only two labelling disagreements occured, which results in 99.61\% inter-rater reliability.

\subsection{Experiment Result Without User-Provided Constraints}{\label{section:experiment_results}}
We evaluated \toolname{} following the scenario described in Section~\ref{section:experiment_setting}. 
The transformation operations for each deprecated API usages are automatically inferred by \toolname{} from their deprecated and updated API signatures. These transformation operations are then applied to the file containing the deprecated API usage, producing the updated file. 
The result of this experiment is shown in the second row of Table~\ref{table:update_evaluation}.


\begin{table}[]
\centering
\caption{\toolname{} experiments results}
\label{table:update_evaluation}
\begin{tabular}{|l|c|c|c|}
\hline
\textbf{Evaluation}                & \textbf{TP} & \textbf{FP} & \textbf{Precision} \\ \hline
Without added constraints & 443         & 71          & 86.19\%            \\ \hline
With added constraints    & 481         & 33          & 93.58\%            \\ \hline
\end{tabular}
\end{table}

From the experiment, \toolname{} achieved a precision of 86.19\%. 
There are 71 updated code that are labelled as false positives, indicating instances of deprecated API usages that were not updated correctly by \toolname{}. One of the main reasons for the presence of false positive results is due to the transformation constraint not being present for context-dependent updates because such a constraint cannot be automatically inferred by \toolname{}.

\subsection{Adding Constraints to \toolname{'s} Inferred DSL}{\label{section:experiment_constraint}}

We further improve the transformation precision of \toolname{} by adding user-specified transformation constraints for context-dependent deprecated API usage updates. Within the evaluation dataset, we found that there are 21 files containing 49 context-dependent deprecated API usages.
Failures in updating these API usages is one of the main contributors for the false positive updates produced by \toolname{}.

While \toolname{} is unable to automatically infer the required transformation constraint from the API signatures, \toolname{} does provide the capability to include such constraints in its transformation.
In particular, \toolname{} accepts transformation constraint in its DSL, as shown in {\tt <transformation_constraint>} non-terminal in Figure~\ref{fig:DSL_Grammar}.


Using this additional input, we conducted a follow-up experiment on \toolname{} using the same test dataset. 
We manually define the required transformation constraints for each context-dependent deprecated API usage migration and add them as an input to \toolname{}.
In total, we manually write 7 transformation constraints.
The result of this follow-up experiment is shown in the third row of Table~\ref{table:update_evaluation}.


Using the additional input constraints, \toolname{} achieves 93.58\% precision, which is a 7.39\% improvement compared to the previous experiment. From the 49 context-dependent deprecated API usages, 38 are updated successfully.
While the addition of the user input constraints improves the overall precision of \toolname{} update result, there remain some false positives. Out of the 514 deprecated API usages, 33 are updated incorrectly.
To understand the reasons for these failures, we conducted a qualitative study, which we will discuss in the next sub-section.

\subsection{\toolname{'s} Failed Migrations}\label{section:failed_migration}
Based on our evaluation of \toolname{}, we see that there are 33 instances of deprecated API usage that are updated incorrectly.
We analyze these cases, aiming to find the reasons of these failed migrations. 
The reasons are:

\begin{enumerate}[nosep,leftmargin=*]
    \item {\em API usage migration involving arithmetic operations}. Deprecated API usage migration may involve arithmetic operations between the updated API usage and other values. This arithmetic operation is not handled by \toolname{}.
    This case is found in the {\tt torch.addcdiv} deprecated API migration, shown in Figure ~\ref{fig:addcdiv_example}.
    In this example, the deprecated API usage (line 1) needs to be replaced into an arithmetic operation between the {\tt input} parameter of the deprecated API usage and a {\tt torch.floor_divide} API invocation (line 2).
    In the future, we can add support for arithmetic operation in \toolname{} to handle this type of API migration.

\begin{figure}
\centering
\scriptsize{
\begin{lstlisting}[xleftmargin=5.0ex,numbers=left,language=diff,sensitive=true,columns=flexible,basicstyle=\ttfamily]
- torch.addcdiv(input, tensor, other)
+ input + torch.floor_divide(tensor, other)
\end{lstlisting}
	\caption{Migration for {\tt torch.addcdiv} deprecated API usage from {\tt Torch} version 1.6.0}\label{fig:addcdiv_example}
}
\end{figure}

    \item {\em API usage migration involving 1:N API mapping}. Currently, \toolname{} only handles API usage migrations with 1:1 and 1:0 API mappings, which are the most common API mapping migration categories of Python deprecated API usage. 
    An example of the 1:N case is found in the {\tt tensorflow.compat.v1.sparse_to_dense} deprecated API migration, where the update requires adding an invocation of {\tt tensorflow.sparse.} {\tt SparseTensor} and changing the API name of {\tt tensorflow.compat.v1.sparse_to_dense} to {\tt tensorflow.sparse_to_dense}. This API migration is illustrated in Figure~\ref{fig:sparsetodense_example}.
    \begin{figure}
    \centering
    \scriptsize{
    \begin{lstlisting}[xleftmargin=5.0ex,numbers=left,language=diff,sensitive=true,columns=flexible,basicstyle=\ttfamily]
- indices_scattered = tf.compat.v1.sparse_to_dense(
-       sparse_indices, tf.cast(tf.shape(shape), 
-       tf.int64), indices_gathered)
+ sparse_tensor = tf.sparse.SparseTensor(sparse_indices, 
+       indices_gathered, tf.cast(tf.shape(shape), 
+       tf.int64))
+ indices_scattered = tf.sparse.to_dense(sparse_tensor)
\end{lstlisting}
    	\caption{Migration for {\tt tensorflow.compat.v1.sparse_ to_dense} deprecated API usage from {\tt Tensorflow} version 2.0.0}\label{fig:sparsetodense_example}
    }
    \end{figure}
    
    In the future, it is possible to extend the operations provided by \toolname{} to include operations adding new API invocations.
    We would also need to enable the automatic transformation inference to infer the required transformations for a 1:N API migration mapping.
    
    \item {\em API usage migration involving value and/or type modifications of the parameter values}. \toolname{} addresses the problem of parameter value or type change through the usage of transformation constraints. However, this approach can be insufficient as some deprecated APIs require the parameter values to be updated, which typically include constructing a new value through mathematical operations or other API invocations.
    An example of this case is found in the {\tt torch.masked_select} deprecated API usage, which is shown in Figure~\ref{fig:parameter_type_example}. In the deprecated version, {\tt torch.masked_select} allows for {\tt IntTensor} as its second positional parameter. In the updated version, the second positional parameter must be of type {\tt BoolTensor}, hence the conversion is done using the {\tt torch.gt} API in line 3.
    In the future, it is possible to add more operations into \toolname{} that would enable modifications of the values of API parameters.

\begin{figure}
\centering
\scriptsize{
\begin{lstlisting}[xleftmargin=5.0ex,numbers=left,language=diff,sensitive=true,columns=flexible,basicstyle=\ttfamily]
    import torch
    def MAPE_torch(pred, mask):
+       mask = torch.gt(true, mask - 0.01)
        pred = torch.masked_select(pred, mask)
\end{lstlisting}
	\caption{Migration for {\tt torch.masked_select} deprecated API usage from {\tt PyTorch} version 1.2.0}\label{fig:parameter_type_example}
}
\end{figure}

    \item {\em API usage migration involving parameter name and position changes of multiple API parameters with the same type}. In this deprecated API migration, the deprecated API parameter names are changed into the new names described in the updated API signature. However, to infer the new names for the deprecated API parameters, \toolname{} makes use of the parameter position and type. If there are multiple parameter name and position changes, \toolname{} may infer an incorrect transformation for the API migration.
    An example of this case is found in the migration of the {\tt torch.stft} deprecated API usage, which is shown in Figure~\ref{fig:stft_DSL}. In this case, there are three API parameters with type {\tt int} that need to be renamed. \toolname{} produces incorrect {\tt rename_parameter} operations for two API parameters, which are {\tt frame_length} parameter that should be renamed into {\tt win_length} and {\tt fft_size} parameter that should be renamed into {\tt n_fft}.
    To mitigate this problem, we can improve the transformation inference done by \toolname{} to not only infer the API parameter transformations based on their type and position, but also through parameter name similarity.
    
    \begin{figure}[t]
	\includegraphics[width=0.90\linewidth]{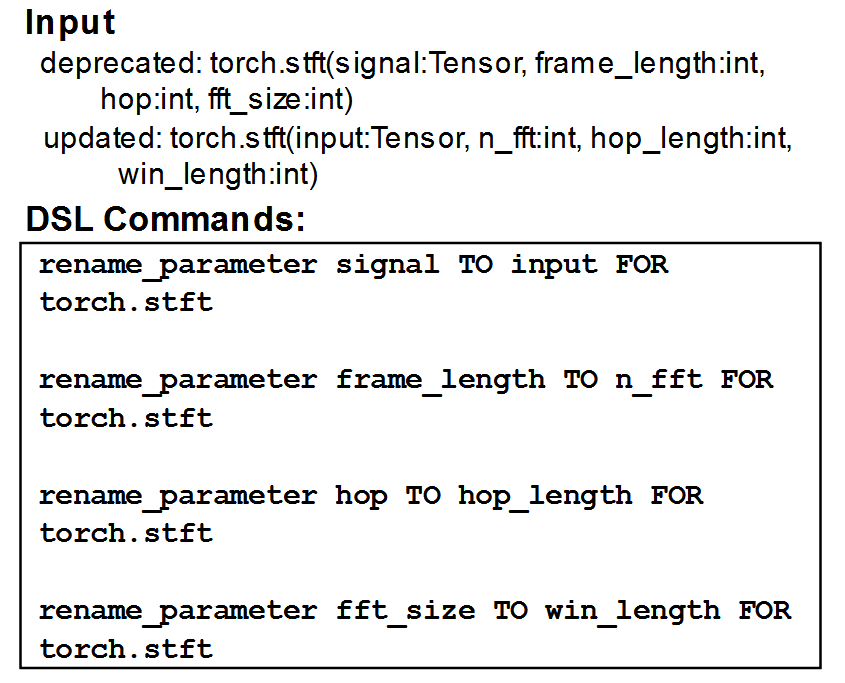}
	\caption{Incorrect DSL commands produced by \toolname{} for {\tt torch.stft} deprecated API usage migration from {\tt PyTorch} version 0.4.1}
	\label{fig:stft_DSL}
    \end{figure}
    

\end{enumerate}

\section{Threats to Validity}\label{section:threats_to_validity}

Threats to {\em external validity} relate to the generalizability of our findings. Our research investigated \numcasestudy{} APIs from three Python machine learning libraries, namely {\tt Scikit-Learn, TensorFlow}, and {\tt PyTorch}. 
These were the most popular machine learning libraries and contain APIs that have been deprecated in the last 2 years. We made this pragmatic choice to address the recent pain points that many developers face. We designed \toolname{} based on the results of our empirical study, which may not hold true for other libraries and APIs. 
There were also cases where \toolname{} is unable to produce correct update, which we discussed in Section~\ref{section:failed_migration}. In the future, we plan to extend our study to cover more recent releases of popular ML libraries -- the landscape of ML libraries may continue to change and other ML libraries may become more popular than the three we investigated. We also plan to address the limitations of \toolname{}, mainly its inability to migrate 1:N API mappings.

Threats to \textit{internal validity} relate to the accuracy and error that might have happen during our analysis. We manually collected the list of deprecated APIs based on the library documentation and changelogs.
From this list, we defined the dimensions and categories of Python deprecated API usage migrations.
It is possible that our list of APIs is incomplete as there might be API deprecations that were not documented properly.
However, as we collected our list of deprecated APIs from the official documentations of each library, we believe that our list is complete enough without any major defect.

Threats to internal validity also apply to the accuracy of the labelling done in the \toolname{} evaluation. As the update result labelling is done by humans, there exists a possibility of human error. To mitigate this problem, each instance of the deprecated API usages is labelled by two different programmers. If any disagreement occurs, a discussion is conducted to achieve a consensus.

Threats to \textit{construct validity} relate to the suitability of the evaluation metric that we chose for \toolname{} evaluation. We used precision as the metric to evaluate the update results of \toolname{} by measuring the percentage of correct update produced by \toolname{} in our test dataset of 258 files. This metric has also been used to evaluate existing tools on automatic update of deprecated APIs~\cite{lamothe_a4, fazzini2019automated, coccievolve, thung_towardsgenerating}. Thus, we believe the threats are minimal.


\section{Related Work}\label{section:related_work}

\textbf{API deprecation}. API deprecation is a topic that has been studied extensively \cite{zhou_API_Deprecation, brito_replacement_message, sawant_deprecation_motivation, Sawant_reaction_deprecation, hora_developers_react ,xavier_impact_api, li2018characterising, bogart_howtobreakapi, Cossette2012SeekingTG}.
Zhou and Walker\cite{zhou_API_Deprecation} created a lightweight framework to detect deprecated API usages in source code examples on the web for Android API usage. Brito et al.~\cite{brito_replacement_message} proposed a recommendation tool to infer replacement messages by mining solutions adopted by developers. This recommendation tool aims to help developers to find deprecated API alternatives. 
Sawant et al.~\cite{sawant_deprecation_motivation, Sawant_reaction_deprecation} explored the reaction of API users towards API deprecations. They found that most developers do not update their API usages due to the cost of update not being worth it. When they do react, they prefer to fix the code by deleting the deprecated API instead of replacing it with the updated API.
A similar study by Hora et al.~\cite{hora_developers_react} explored the impact of API evolution and found that API changes can affect the whole ecosytem. They also found that client developers need some time to discover and apply the new APIs, while the majority does not react at all. Xavier et al.\cite{xavier_impact_api} found that the frequency of API deprecations increase over time and systems that are more popular have higher frequency of deprecations. Li et al.~\cite{li2018characterising} characterize deprecated Android APIs and found that usage of deprecated APIs are mostly found in popular libraries.
Our study is focused on the API deprecation of Python machine learning libraries. We manually collected a list of \numcasestudy{} deprecated APIs from three popular Python machine learning libraries, {\tt Scikit-learn}, {\tt PyTorch}, and {\tt Tensorflow}. We investigated the common patterns in the API migration of these deprecated APIs.

\textbf{API migration}. Multiple works and studies on API migration have been done~\cite{bogart_howtobreakapi, Cossette2012SeekingTG, hora_apiwave, li_webapievolution, dig_roleofrefactoring, xia_howandroidevelopers}. Bogart et al.~\cite{bogart_howtobreakapi} found that different programming ecosystems have different expectations towards API breaking changes. 
Cossette and Walker~\cite{Cossette2012SeekingTG} conducted a retroactive study on API incompatibilities between several versions of Java libraries. From their manual analysis, they found that the majority of API changes cannot be automatically transformed without severe restrictions and that the majority of changes to an API were not documented properly.
The tool apiwave~\cite{hora_apiwave} keeps track of API popularity and migration of major frameworks and libraries from top 650 GitHub Java projects.
Li et al.~\cite{li_webapievolution} conducted an empirical study on web API evolution by analyzing the evolution of five popular web APIs. They found that web APIs evolve in limited patterns and that most of the API changes in web APIs involve refactoring.
Dig and Johnson~\cite{dig_roleofrefactoring} study the requirements for API migration tools by analyzing the API changes of three frameworks and one library in Java. They found that from their case studies, more than 80\% of the API changes are refactoring, indicating an important role of refactoring in API migration. Xia et al.~\cite{xia_howandroidevelopers} performs large-scale study on the current practice of handling evolution-induced compatibility issues in Android apps. They proposed RAPID, an automated tool to determine whether a compatibility issue has been addressed or not through utilization of static analysis and machine learning techniques.
Our study is focused on the API migration of Python machine learning deprecated API usages. We defined the dimensions of API migration for Python machine learning deprecated API usages using the case study of \numcasestudy{} deprecated APIs.

\textbf{Program transformation}. Works on program transformation has also been done~\cite{LASE, rolim_refazer, lawall2018coccinelle, brunel2009foundation, kang2019semantic, fazzini2019automated, coccievolve, thung_towardsgenerating, meng_sydit, meng_sydit2, lamothe_a4, lucas_spinfer}. LASE~\cite{LASE} is an Eclipse plug-in that provides an example-based program transformation.
LASE is able to create a context-aware edit script from code examples, and uses the script to transform the code.
Sydit is a program transformation tool that helps developers in systematic editing task involving similar changes in multiple places. Given a source and target method, Sydit learns and applies systematic edits which results are displayed to the programmer to review and approve.
REFAZER~\cite{rolim_refazer} was proposed by Rolim et al. to automatically learn program transformations through observation of code edits performed by developers. REFAZER synthesize program transformations for Python code by leveraging inductive programming using input-output examples. Coccinelle\cite{lawall2018coccinelle, brunel2009foundation}, a program matching and transformation tool for C language, allows program developers to write code manipulation rules in terms of code structure. Transformation in Coccinelle is expressed in the form of semantic patch language (SmPL), which has a syntax similar to a code diff. A port of Coccinelle for Java language, called Coccinelle4J was also proposed by Kang et al.~\cite{kang2019semantic}. 
SPINFER~\cite{lucas_spinfer} is capable of automatically inferring Coccinelle semantic patches from existing code change examples for the Linux kernel. SPINFER considers similar code fragments and control flows in the changes to identify the change patterns.

A4~\cite{lamothe_a4} is an approach to assist developers with Android API migration by leveraging source code examples to learn the API migration patterns. A4 can either automatically migrate API calls with little to no extra modifications or provide a guidance to assist with the migration.
More recently, Fazzini et al.~\cite{fazzini2019automated} created AppEvolve which transforms the usage of deprecated Android APIs into backward-compatible updated code by learning from code examples. Haryono et al.~\cite{coccievolve} presented a new tool for automatic Android API update called CocciEvolve which is
built on Coccinelle4J. CocciEvolve shows an improvement from AppEvolve in the form of readable transformations and its capability to produce working updates with only a single update example.
Thung et al.~\cite{thung_towardsgenerating} proposed NEAT, a tool to generate transformation rules that can assist developers in deprecated API replacement without code example for Android APIs. 
NEAT uses the signature graph from the source code of the API library to convert between one type and another.
In our work, we proposed \toolname{}, which automates the update of Python machine learning deprecated API usages. \toolname{} does not require any code example to provide its transformation, but rather infers the required transformations through the difference between the deprecated and updated API signatures and presents these transformations in the form of a DSL. \toolname{} is built for the Python programming language, bringing a set of difficulties previously unseen in other programming languages, 
such as positional and keyword parameters, default parameter values, usage of whitespaces as separator, and dynamic typing which allows a variable to have different types during the program execution.

\section{Conclusion and Future Work}\label{section:conclusion_future_work}
We conducted an empirical study to learn the dimensions of the API migrations required by Python deprecated APIs. 
We manually collected a list of \numcasestudy{} APIs and their updated API signatures from three popular Python machine learning libraries: {\tt Scikit-Learn, TensorFlow}, and {\tt PyTorch}.
Using manual thematic analysis, we found three dimensions in the deprecated API migrations: 
 update operation, API mapping, and context dependency. 

Based on findings in our empirical study, we created \toolname{}, an automated tool to update the usage of Python deprecated APIs. \toolname{} 
automatically infers the transformation operations to migrate and update the usage of a deprecated API. Given the API signatures and the file containing deprecated API usages, \toolname{} automatically lists the  transformations in the form of a DSL and updates the deprecated API usage within the file.
We evaluated \toolname{} using a dataset of 258 files containing 514 API usages from 66 different APIs. 
\toolname{} achieved 86.19\% precision. We further improve \toolname{} by adding transformation constraints input, allowing it to process the update for context-dependent deprecated API usage update. With this improvement, \toolname{} achieved 93.58\% precision, a 7.39\% precision improvement from experiment without user-added constraint.


For future work, we plan to extend our study to other popular Python libraries, including non-machine learning libraries. By increasing the scope of the case study, we may encounter more variety of deprecated APIs and their migrations.
We also plan to address the current shortcomings of \toolname{}, namely its inability to create 1:N API mapping migration updates, etc.




\balance

\bibliography{references}
\bibliographystyle{IEEEtran}

\end{document}